\documentstyle[epsf,referee]{l-aa}
\begin{document}
%______________________________________________________________________
\thesaurus{06(06.13.1; others to be inserted latter)}
\title{Axisymmetric mean field dynamos
with dynamic and algebraic $\alpha$--quenchings}
%______________________________________________________________________
\author{Eurico Covas\thanks{e-mail: E.O.Covas@qmw.ac.uk}\inst{1}
\and    Reza Tavakol\thanks{e-mail: reza@maths.qmw.ac.uk}\inst{1}
\and    Andrew Tworkowski\thanks{e-mail: A.S.Tworkowski@qmw.ac.uk}\inst{1}
\and    Axel Brandenburg\thanks{e-mail: Axel.Brandenburg@ncl.ac.uk}\inst{2}}
%______________________________________________________________________
\institute{Astronomy Unit, School of Mathematical Sciences,
Queen Mary and Westfield College, Mile End Road, London E1 4NS, UK
\and Department of Mathematics, University of Newcastle
upon Tyne NE1 7RU, UK}
%______________________________________________________________________
\date{Received ~~ ; accepted ~~ }
\offprints{Eurico Covas}
\maketitle
\markboth
{Covas et al.: Dynamos with dynamic and algebraic $\alpha$--quenchings}
{Covas et al.: Dynamos with dynamic and algebraic $\alpha$--quenchings}
%______________________________________________________________________
\begin{abstract}
%______________________________________________________________________
We study axisymmetric
mean field spherical and spherical shell dynamo models,
with both dynamic and algebraic $\alpha$--quenchings.
Our results show that there are qualitative as well as
quantitative differences and similarities between these models.
Regarding similarities, both groups of models
exhibit symmetric, antisymmetric and mixed modes of behaviour.
As regards differences, the important
feature in the full sphere models
is the occurrence of chaotic behaviour
in the algebraic $\alpha$--quenching models.
For the spherical shell models with dynamic $\alpha$
the main features include the
possibility of multi-attractor regimes with
final state sensitivity with
respect to small changes in the magnitude of $\alpha$
and the initial parity.

We find the effect of introducing a
dynamic $\alpha$ is likely
to be complicated and depend on the region of the
parameter space considered, rather than a uniform
change towards simplicity or complexity.

\end{abstract}
%______________________________________________________________________
\section{Introduction}
\label{intro}
%______________________________________________________________________
Axisymmetric mean field dynamo models have been extensively studied as
a  possible way  of understanding  some aspects  of solar and  stellar
dynamos   (see for example, Brandenburg et al.\ 1989a,b; Moss   et
al.\ 1991; Tavakol et al.\ 1995). In  most such studies the
nonlinearity is introduced through an algebraic (as opposed to a
dynamic) form of $\alpha$--quenching.  These studies  have produced
a number of  novel modes  of  behaviour in spherical and   spherical
shell dynamo models, which include periodic,  quasiperiodic and
chaotic solutions, with the latter mode of  behaviour  previously
observed  only in  spherical shell (Tavakol et al.\ 1995), torus
(Brooke \& Moss 1994), and accretion disc dynamos
(Torkelsson \& Brandenburg 1994, 1995).  In addition it has recently been
shown that spherical shell models
are capable of producing  various forms of intermittent
type behaviour (Tworkowski  et al.\ 1997).
Similar results have also been obtained in  torus
models by Brooke \& Moss (1995), Brooke (1997), Brooke et al.\ (1997).
Such intermittent behaviour could be of relevance in
understanding some  of the intermediate time scale variability
observed in  the output of the Sun (Eddy 1975; Ribes \& Nesme-Ribes
1993) and  stars (Baliunas \& Vaughan 1985).

An important  point  regarding the  $\alpha$--quenching terms, usually
employed in such models, is  that they are  approximate  in nature. This
is a direct consequence of the fact that small scale
turbulent effects cannot
be prescribed precisely and parametrisations are
always required in order
to estimate second order correlations. As
a   result the question arises as to whether the modes  of behaviour
observed in such studies  are  in some sense  a consequence of the
particular forms of $\alpha$--quenching  employed.  The question   of
robustness  of such dynamo  models  with respect to reasonable
changes  in   the functional form  of $\alpha$--quenching was considered in
Tavakol et al.\ (1995),  where it was shown that  the dynamics was
qualitatively    robust with respect to   such changes.  The    forms
of    $\alpha$ considered  there  were, however, algebraic. One
potential    problem with the algebraic  forms of
$\alpha$--quenching is that they act instantaneously and this may have a
bearing on the occurrence of  the more complicated modes of behaviour,
such as chaos and intermittency,
observed in such models.

A possible way of remedying this latter shortcoming is to relax
the instantaneous feature by employing  a dynamic
(explicitly time  dependent) $\alpha$ effect.  The underlying physics
of this type of effect has been  given by Kleeorin \& Ruzmaikin (1982)
and Zeldovich et al.\ (1983) (see also  Kleeorin et al.\ (1995)),
where  the  existence  of    non-instantaneous quenching was shown to be a
consequence of the fact that the magnetic helicity is conserved in the
absence of diffusion or boundary effects.

Truncated one-dimensional models with dynamic $\alpha$ effect
have recently been investigated by Schmalz \& Stix
(1991), Covas et al.\ (1997a,b) and Covas \& Tavakol (1997).
There is also
a two dimensional truncated analogue of such results
studied by Schlichenmaier \& Stix (1995).

Our aim here is to make
a detailed study of the dynamics of
axisymmetric mean field dynamo models
with a dynamic $\alpha$--quenching in order to find out
how they compare with models involving
algebraic $\alpha$--quenching and in particular
whether the novel
features discovered in such settings, such as chaotic
and intermittent-type behaviour survive as
$\alpha$--quenching is made dynamic.
To make this study comparative, it was necessary to
extend the previous results of Brandenburg et al.\ (1989a,b) and
Tavakol et al.\ (1995)
which employed an algebraic $\alpha$--quenching.

The structure of the paper is as follows. In section 2 we
briefly introduce the equations for the axisymmetric
mean field dynamo models with both dynamic and algebraic $\alpha$--quenchings.
Section
3 contains our results for each case, with sections 3.1 and 3.2 summarising
our results for spherical and spherical shell dynamo models respectively.
In section 4 we briefly discuss the occurrence of multiple attractor regimes
for these models and in section 5 we report on the
presence of intermittent types of behaviour
in the shell models. Finally we present in section 6 our conclusions.
%______________________________________________________________________
\section{The model}
%______________________________________________________________________
The standard mean field dynamo equation (cf. Krause \& R\"adler 1980) is
of the form
%______________________________________________________________________
\begin{equation}
\label{dynamo}
\frac{\partial \vec{B}}{\partial t}=\nabla \times \left( \vec{u}
\times \vec{B} + \alpha \vec{B} - \eta_t \nabla \times \vec {B} \right),
\end{equation}
%______________________________________________________________________
where $\vec{B}$ and $\vec{u}$ are the mean
magnetic field and the mean velocity respectively
and $\eta_t$ is the turbulent magnetic diffusivity.
For simplicity, and to facilitate comparison with previous work, $\alpha$
and $\eta_t$ are assumed to be scalars.
The magnitudes of the $\alpha$ and $\omega$ effects
are given by the dynamo parameters
$C_{\alpha}$ and $C_{\omega}$.
Even though our main aim here is to study the effects of making
$\alpha$ dynamical, nevertheless, for the sake of comparison,
it is necessary to consider analogous models with
algebraic $\alpha$--quenching.
We therefore consider both cases.

In the algebraic case we take the functional form of
algebraic $\alpha$--quenching to be the usual one, namely
%______________________________________________________________________
\begin{equation}\label{alpha_a}
\alpha_a=\frac{\alpha_0 \cos \theta}{1+ \vec{B}^2},
\label{aquench}
\end{equation}
%______________________________________________________________________
an expression that has been adopted in numerous studies since Jepps (1975).
In the dynamical case, $\alpha$ can, according to Zeldovich
et al. (1983) and Kleeorin \& Ruzmaikin (1982) (see
also Kleeorin et al.\ 1995),
be divided into a
hydrodynamic and a magnetic part, thus
%______________________________________________________________________
\begin{equation}
\label{allalpha}
\alpha=\alpha_h+\alpha_m,
\end{equation}
%______________________________________________________________________
\label{alpha_full}
where $\alpha_h = \alpha_a$ is given by Eq.~(\ref{aquench}) and
the magnetic part satisfies an explicitly time dependent diffusion type
equation with a nonlinear forcing in the form
%______________________________________________________________________
\begin{equation}
\label{alpha_d}
\frac{\partial\alpha_m}{\partial t}=\frac{1}{\mu_0 \rho}(\vec{J}\cdot\vec{B}-\frac{\alpha\vec{B}^2}{\eta_t})
+\nu_{\alpha}\nabla^2\alpha_m,
\end{equation}
%______________________________________________________________________
where $\nu_{\alpha}$ is a physical constant parameter taken to be
half times the value of $\eta_t$,
$\rho$ the density of the medium, $\mu_0$ the induction constant and in the
$\alpha\vec{B}^2$ term the full $\alpha$ from Eq. (\ref{allalpha}) is
used. This equation is slightly different from that given by
Kleeorin \& Ruzmaikin (1982), where the damping term has the form
$-\alpha_m/\tau_\alpha$ (where $\tau_\alpha$ is a typical
relaxation time) instead of $+\nu_{\alpha}\nabla^2\alpha_m$.
For a detailed comparison between different variations of this
equation see Covas et al.\ (1997a,b).

The models we shall consider here are in the forms of sphere and
spherical shells, with the outer boundary in both cases
being denoted by $R$ and in the case
of spherical shell models, the fractional radius of the inner
boundary of the shell being denoted by
$r_0$.
As is customary, we shall in the following, discuss the behaviour of
the dynamos considered by monitoring the total magnetic energy,
$E={1\over2\mu_0}\int\vec{B}^2dV$,  in $r\leq R$. We split $E$ into
two parts,
$E=E^{(A)}+E^{(S)}$, where $E^{(A)}$ and $E^{(S)}$ are respectively
the energies of those parts of the field whose toroidal field is
antisymmetric and symmetric about the equator.
The overall parity $P$ given by $P=(E^{(S)}-E^{(A)})/E$. In this way,
$P=-1$ denotes a antisymmetric (dipole-like) pure parity solution and
$P=+1$ a symmetric (quadrupole-like) pure parity solution.

The physically interesting case is when $\nu_{\alpha}<\eta_t$
corresponding to an adjustment time of the magnetic $\alpha$ effect
that is longer than the magnetic diffusion time (cf. Zeldovich et
al. 1983). For $\vec{B}$ we assume vacuum boundary conditions and for
$\alpha_m$ we use
%______________________________________________________________________
\begin{equation}
\alpha_m=0 \quad \mbox{on} \quad r=R.
\end{equation}
%______________________________________________________________________

For the numerical studies reported in the following sections,
we used a modified version of the axisymmetric
dynamo code of Brandenburg et al.\ (1989a).
We considered full sphere and spherical shell models
and used a grid size of
$41\,\times\,81$ mesh points and employed single precision arithmetic,
ie 4 byte floating point real numbers. To test the robustness of
the  code we verified that no qualitative changes were produced by
employing a finer grid, different temporal step length (we used a step
length of $10^{-4}R^2/\eta_t$ in  the results presented in this paper)
and/or by increasing the machine precision.
Our results are presented in the following sections.
%______________________________________________________________________
\section{Results}
%______________________________________________________________________
We studied the two cases, with algebraic and dynamic $\alpha$--quenchings
separately. For the dynamical case
we solved equations (\ref{dynamo}) and (\ref{alpha_d})
and for the case with the algebraic $\alpha$--quenching
we solved equations (\ref{dynamo}) and  (\ref{alpha_a}).
The latter extends previous studies by
Brandenburg et al.\ (1989a,b) and Tavakol et al.\ (1995).

We should emphasise that all our conclusions
presented here (such as transitions between different
modes of behaviour) are subject to the necessarily finite resolution of
the parameter space chosen. In other words, it is possible that
we have missed out certain types of solutions either within our
relatively coarse parameter mesh, or outside.

Furthermore, an important point regarding the comparison
of these models is that the parameters $C_{\alpha}$ and
$C_{\omega}$ do not play identical roles in these
models and as a result the comparison of these
models, using these parameters, is problematic.
Since our main aim is to study the types of possible behaviour
allowable in the supercritical regimes, we therefore took the
following strategy. Firstly, we chose $C_{\omega}$ values which
are effectively the highest values numerically allowed
by our code, which turned out to be $-10^4$ in the dynamic case
and $-10^5$ in the algebraic case. However, we also
studied the algebraic case at $C_{\omega}=-10^4$, which is the
same value used also in the dynamic case.

The notation used in the figures throughout the paper is as follows:
``A'' represents fixed antisymmetric solutions about the equator,
``S'' fixed symmetric,
``OM'' oscillatory mixed, ``M'' non-oscillatory mixed
and ``C'' chaotic solutions.
We also introduce unstable branches in the pictures when
these are distinct enough.
%--------------------------------------------------------------
\subsection{Spherical dynamo models}
\begin{figure}
\centerline{
\def\epsfsize#1#2{0.5#1}\epsffile{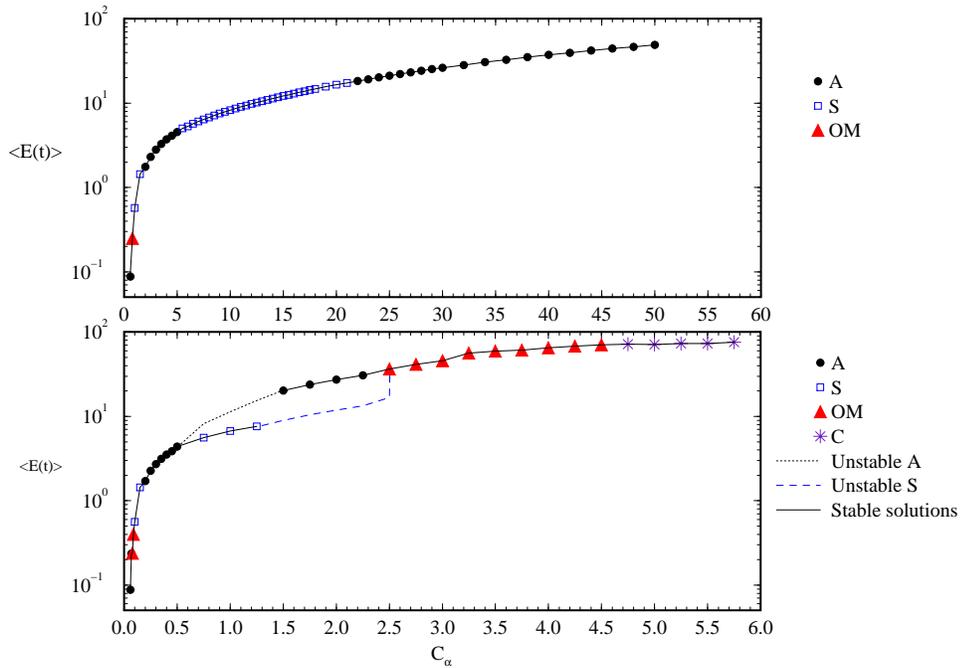}}
\caption[]{\label{fullstatic}
Energy diagram for the full sphere ($r_0=0$) with
algebraic $\alpha$.
The value of $C_{\omega}$ used was $-10^4$
in the top panel and $-10^{5}$ in the bottom one.}
\end{figure}

\begin{figure}
%\picplace{7cm}
\centerline{\def\epsfsize#1#2{0.5#1}\epsffile{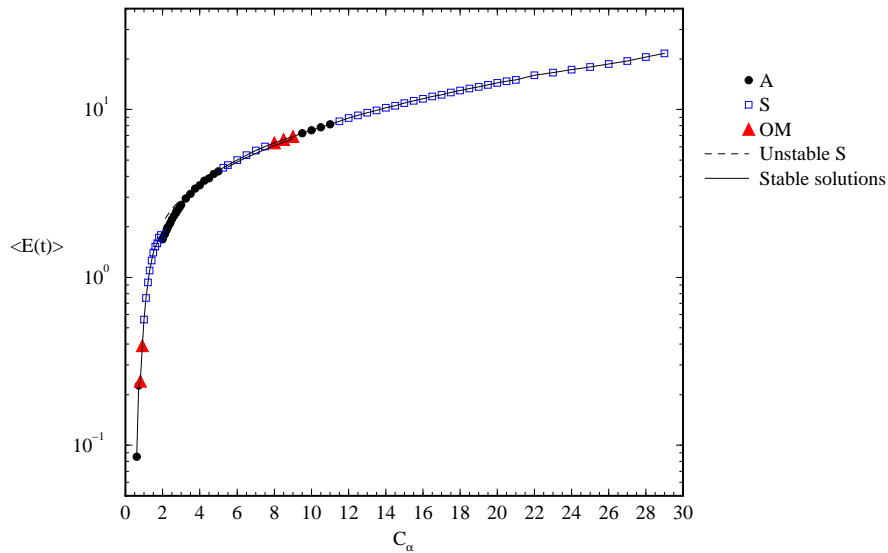}}
\caption[]{\label{full}
Energy diagram for the full sphere ($r_0=0$) with dynamic $\alpha$.
The value of $C_{\omega}$ used was $-10^{4}$.}
\end{figure}

%--------------------------------------------------------------
The results of our computations for the spherical case with algebraic
$\alpha$--quenching are summarised in Fig.\ \ref{fullstatic},
in which the top and the bottom panels are
for $C_{\omega}$ values $-10^4$ and $-10^5$ respectively.
An important new feature of these
results is the presence of chaotic behaviour in the
case where $C_{\omega}=-10^5$. As far as we are aware,
this is the first time chaos has been
observed in a full sphere (as opposed to spherical shell) model of this type.
The sequence of transitions in these two cases
are different at high $C_{\alpha}$ values.
%______________________________________________________________________
%______________________________________________________________________
The corresponding results with dynamic $\alpha$ are presented in
Fig.\ \ref{full}. The sequence of transitions in this case
is antisymmetric $\rightarrow$ symmetric $\rightarrow$ mixed
$\rightarrow$ antisymmetric $\rightarrow$ symmetric,
which is more complicated than the antisymetric $\rightarrow$ symmetric
$\rightarrow$ antisymetric sequence for the algebraic case.

Comparing Figs.\ \ref{fullstatic} and 2
we observe both similarities and differences
between them.  Qualitative similarities include the occurrence of
symmetric, antisymmetric and mixed modes and the absence of
intermittent types of behaviour. The differences lie in the
details of transitions and more importantly, the
way the introduction of dynamic $\alpha$
appears to remove the possibility of the occurrence of
chaotic behaviour in such models (at least when comparing with the
algebraic case for $C_\omega=-10^5$), whilst at the same time
increasing the likelihood of the occurrence of the oscillatory
mixed modes at $C_\omega=-10^4$. We should note that the
reason why such chaotic behaviour was not discovered in
Tavakol et al.\ (1995) was that these authors only considered
a value of $-10^4$ for $C_{\omega}$.
Also important is the stronger increase in the time averaged energy
of the solutions for the algebraic case as $C_{\alpha}$ increases.
As we shall see latter on this feature is quite frequent
in the algebraic models. To summarise, the
effect of making $\alpha$ dynamical is therefore substantial,
with both qualitative and quantitative changes.
%______________________________________________________________________
%______________________________________________________________________
\subsection{Spherical shell dynamo models}
%--------------------------------------------------------------
In the case of the shell dynamos,
we assumed in both the dynamic and algebraic cases
that the $\alpha^2\omega$ dynamos were situated in a conducting
fluid with a vacuum outer boundary condition. The inner boundary
condition was assumed to be a superposition of perfectly conducting
and penetrative boundary conditions in the forms (Tavakol et al.\ 1995):
%______________________________________________________________________
\begin{equation}
(1-F)a+F \left( {{\partial a} \over {\partial r}}-{{a}
\over {\delta}} \right )=0,
\end{equation}
%______________________________________________________________________
and
%______________________________________________________________________
\begin{equation}
(1-F) \left ({{1} \over {r}} {{\partial (rb)}
\over {\partial r}}-\alpha {{1} \over{r}} {{\partial (ra)}
\over {\partial r}} \right )+F
\left ({{\partial b} \over {\partial r}}-{{b} \over{\delta}} \right)=0.
\end{equation}
%______________________________________________________________________
In this way the
boundary conditions may be changed by varying $F$, with $F=0$
corresponding to the perfect conductor case and $F=1$
to the case where the magnetic field goes to zero
at some distance $\delta$ below the inner boundary (Brandenburg et al.\ 1992).
In our numerical computations we considered shells of different
thickness, quantified by the parameter $r_0$ (we took
$r_0 =0.2, 0.5, 0.7$ in units of the outer shell radius)
and in each case took different
values of $F$ ($F=0, 0.5, 1$).  In this way
we were able to study the robustness of spherical shell dynamo models
with respect to changes in both thickness and boundary conditions.

%______________________________________________________________________
\begin{figure}
%\picplace{7cm}
\centerline{\def\epsfsize#1#2{0.5#1}\epsffile{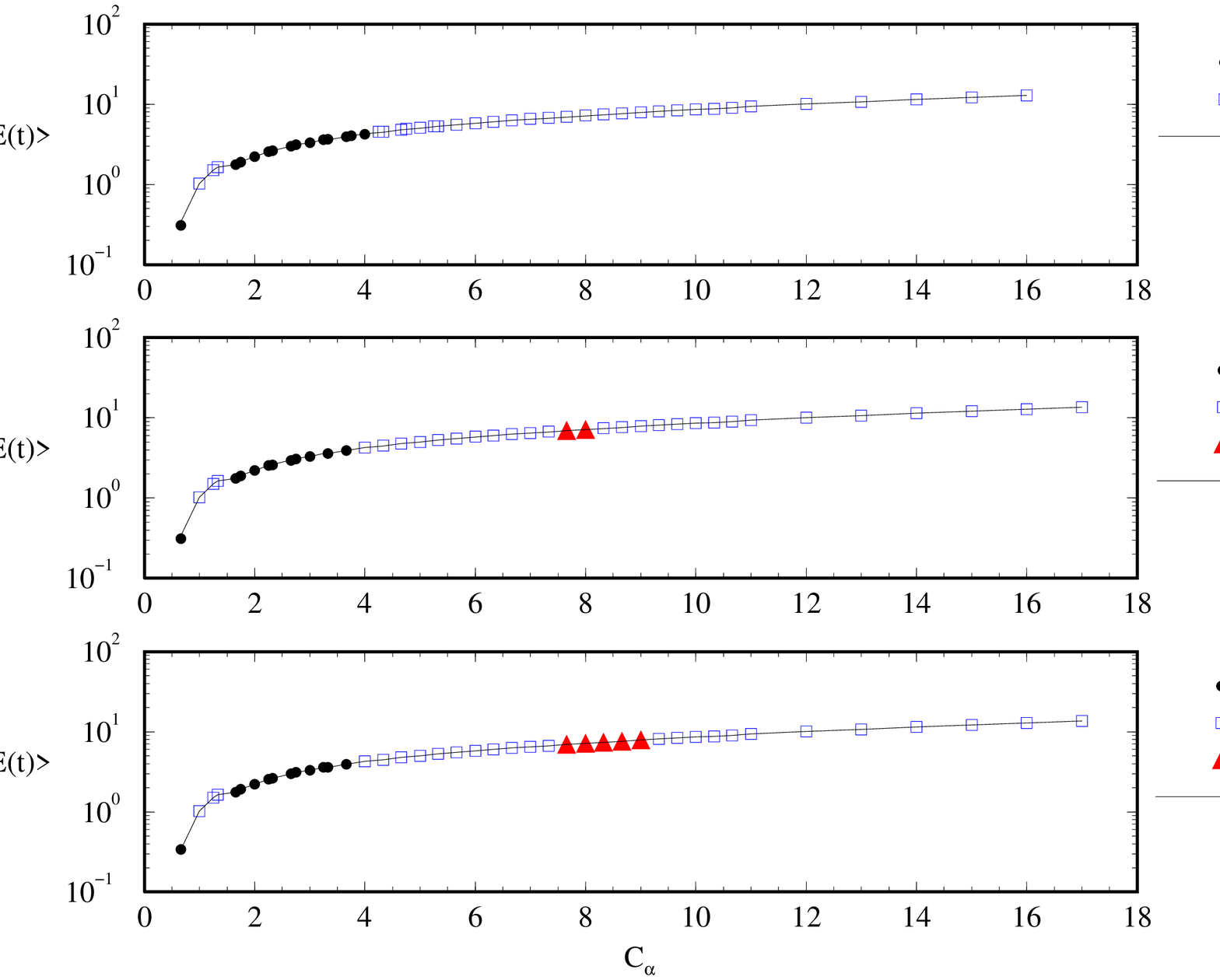}}
\caption[]{\label{thick}
Energy diagrams for the thick shell dynamo $(r_0=0.2)$  with dynamic $\alpha$ and
with $F=0, 0.5$ and $1$.The value of $C_{\omega}$ used was $-10^{4}$. }
\end{figure}
%______________________________________________________________________

%______________________________________________________________________
\begin{figure}
%\picplace{7cm}
\centerline{\def\epsfsize#1#2{0.5#1}\epsffile{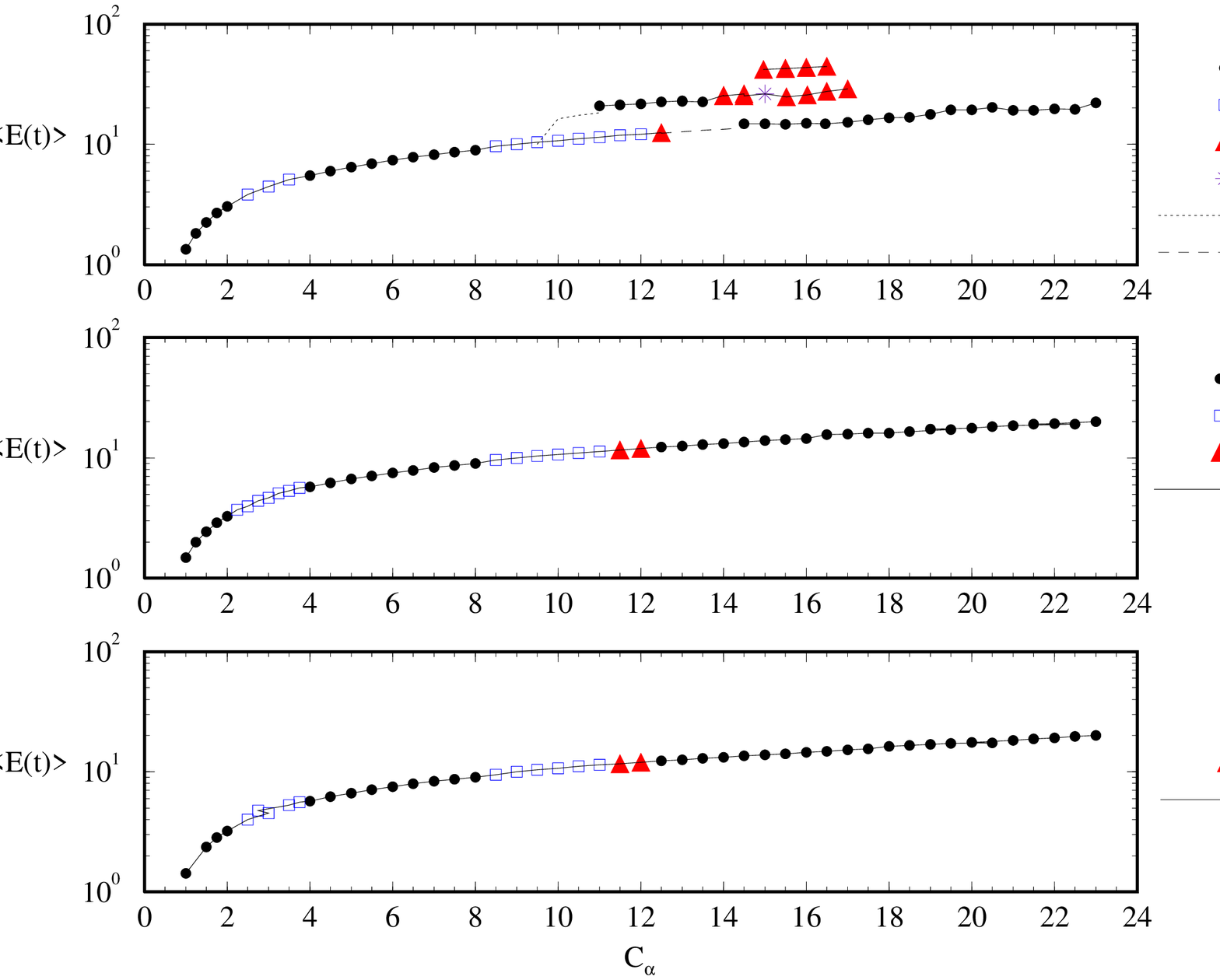}}
\caption[]{\label{medium}
Energy diagrams for a medium shell dynamo $(r_0=0.5)$  with dynamic $\alpha$
and with $F=0, 0.5$ and $1$.
The value of $C_{\omega}$ used was $-10^{4}$.}
\end{figure}
%______________________________________________________________________

%______________________________________________________________________
\begin{figure}
%\picplace{7cm}
\centerline{\def\epsfsize#1#2{0.5#1}\epsffile{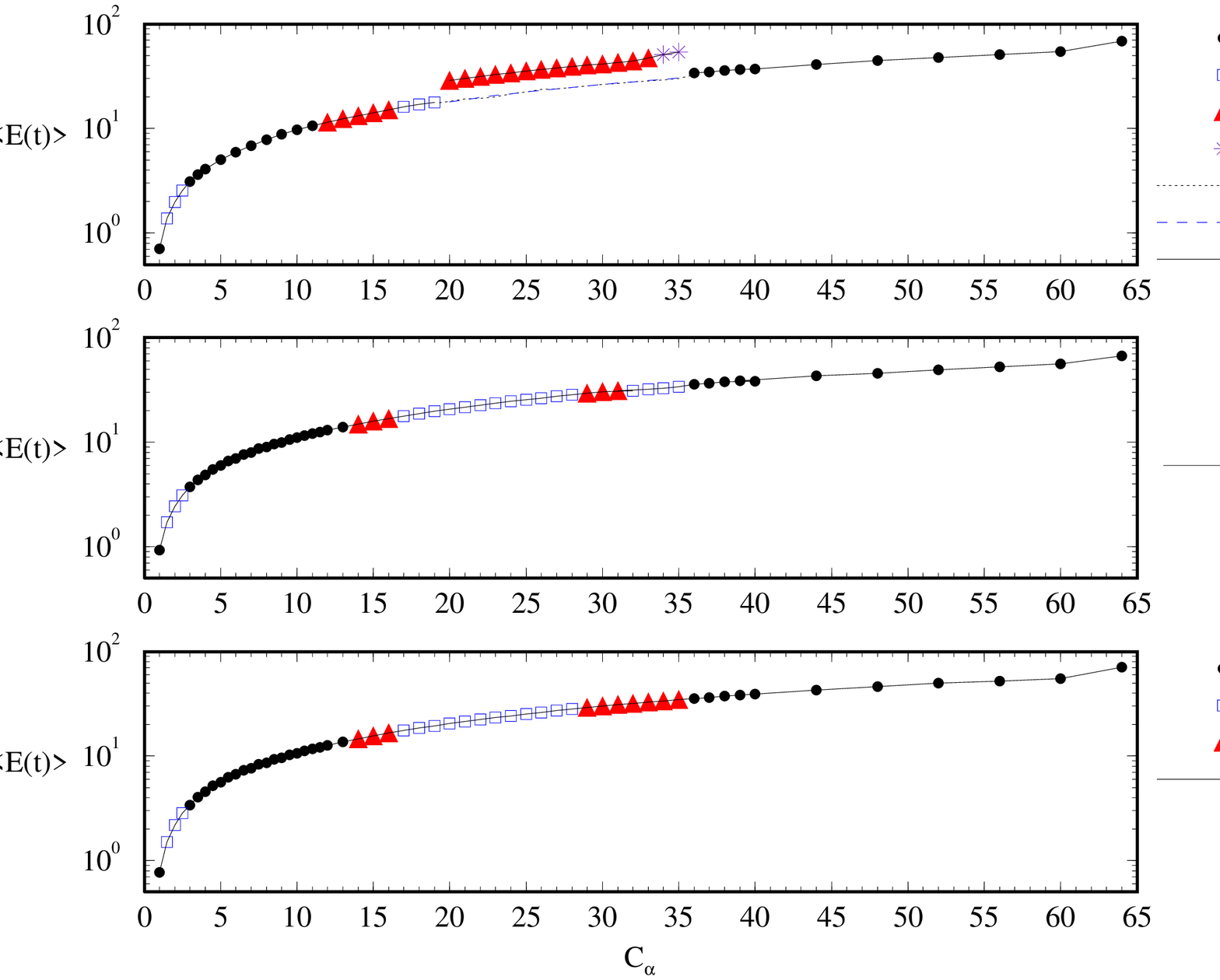}}
\caption[]{\label{thin}
Energy diagrams for a thin shell dynamo $(r_0=0.7)$  with dynamic $\alpha$
and with $F=0, 0.5$ and $1$.
The value of $C_{\omega}$ used was $-10^{4}$.}
\end{figure}
%______________________________________________________________________

%______________________________________________________________________
\begin{figure}
%\picplace{7cm}
\centerline{\def\epsfsize#1#2{0.5#1}\epsffile{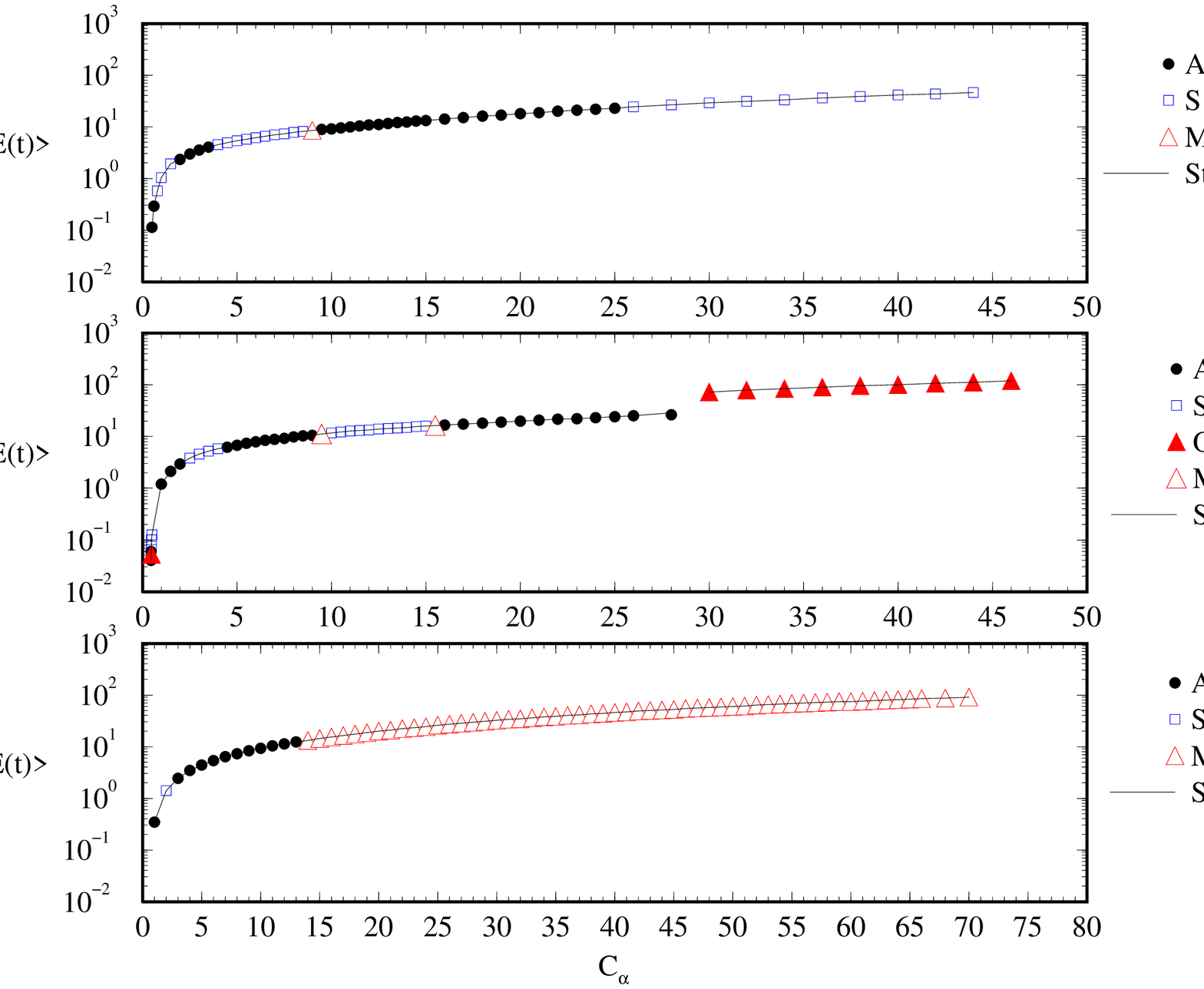}}
\caption[]{\label{four_algebraic}
Energy diagram for algebraic $\alpha$.
The panels from top to bottom correspond to
$r_0$ values $0.2$, $0.5$ and $0.7$ respectively.
The value of $C_{\omega}$ used was $-10^{4}$
and $F=0$.}
\end{figure}
%______________________________________________________________________

In the dynamical case,  the results of our computations
in shells of different thickness, and in each case with
the three values of $F$, are depicted in Figs.\
\ref{thick}--\ref{thin} respectively.
As can be seen, the behaviour of the thicker shell dynamo model ($r_0=0.2$)
consists of an initial sequence of
antisymmetric $\rightarrow$ symmetric $\rightarrow$ antisymmetric
$\rightarrow$ symmetric
sequence, with the $F>0$ models having their asymptotic symmetric states
being slightly interrupted by
a mixed regime. An important feature of these
thicker shell models
is that the symmetric
modes appear to be more likely in the $C_{\alpha}$ measure sense.
The behaviour of the thinner shells ($r_0=0.5$ and $r_0=0.7$)
seem more varied dynamically. Important new features here are
the occurrence
of multiple attractors at intermediate values of $C_{\alpha}$,
for models with $F=0$, as well as intermittent and chaotic
modes of behaviour.
Also these models seem to have an antisymmetric
asymptotic state,  with the antisymmetric modes dominating
in the $C_{\alpha}$ measure sense, in contrast
to the symmetric states in the
$r_0=0.2$ case.
The comparison of these figures also shows that, overall,
thicker shells are more robust to changes in $F$,
at least for $F>0$, which seems to indicate that
$F=0$ models (perfect conductor) are rather special in this case,
in the $F$ measure sense.

In the above we have compared models with dynamic and algebraic $\alpha$
for different values of $C_{\omega}$, because we felt $C_{\omega}$
should be as large as numerically permissible. However, we also
considered models with algebraic $\alpha$-quenching and the same
value of $C_{\omega}$ as in the dynamic case.
Those results are shown in Fig.\ \ref{four_algebraic} and, as can be seen,
there is no evidence for complicated or chaotic behaviour.
We also note that most of the mixed mode solutions possess
non-oscillatory mixed parity and are therefore
in this sense simpler.
We add that we confined ourselves to the $F=0$ case
in this picture, since this produced the most interesting
dynamical behaviour in the dynamic $\alpha$ case, as
we shall see below.

\begin{figure}
\centerline{
\def\epsfsize#1#2{0.5#1}\epsffile{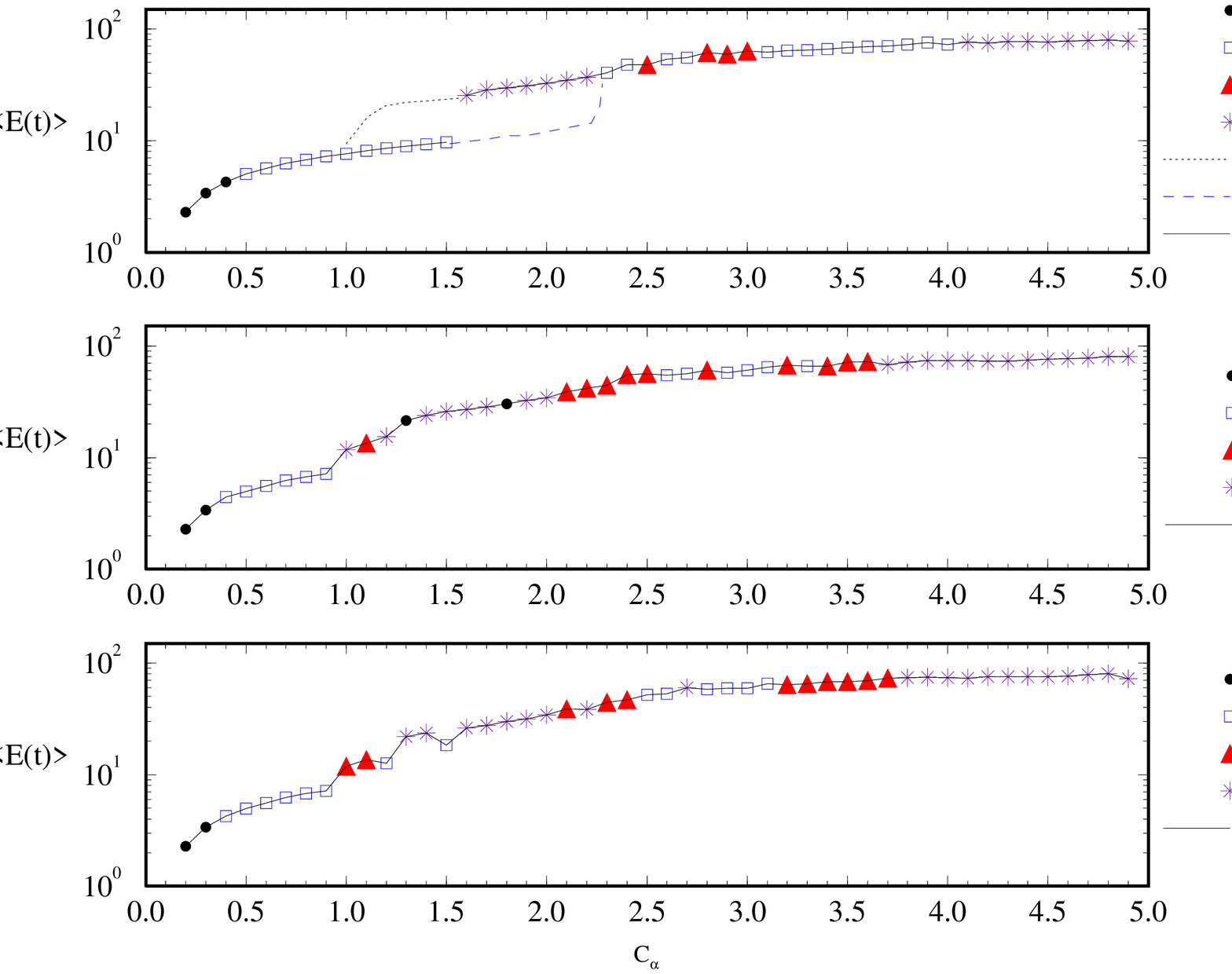}}
\caption[]{\label{thickstatic}
Energy diagrams for a thin shell with algebraic $\alpha$
dynamo $(r_0=0.2)$ and with $F=0, 0.5$ and $1$. The value of $C_{\omega}$ used was $-10^{5}$. }
\end{figure}

\begin{figure}
\centerline{
\def\epsfsize#1#2{0.5#1}\epsffile{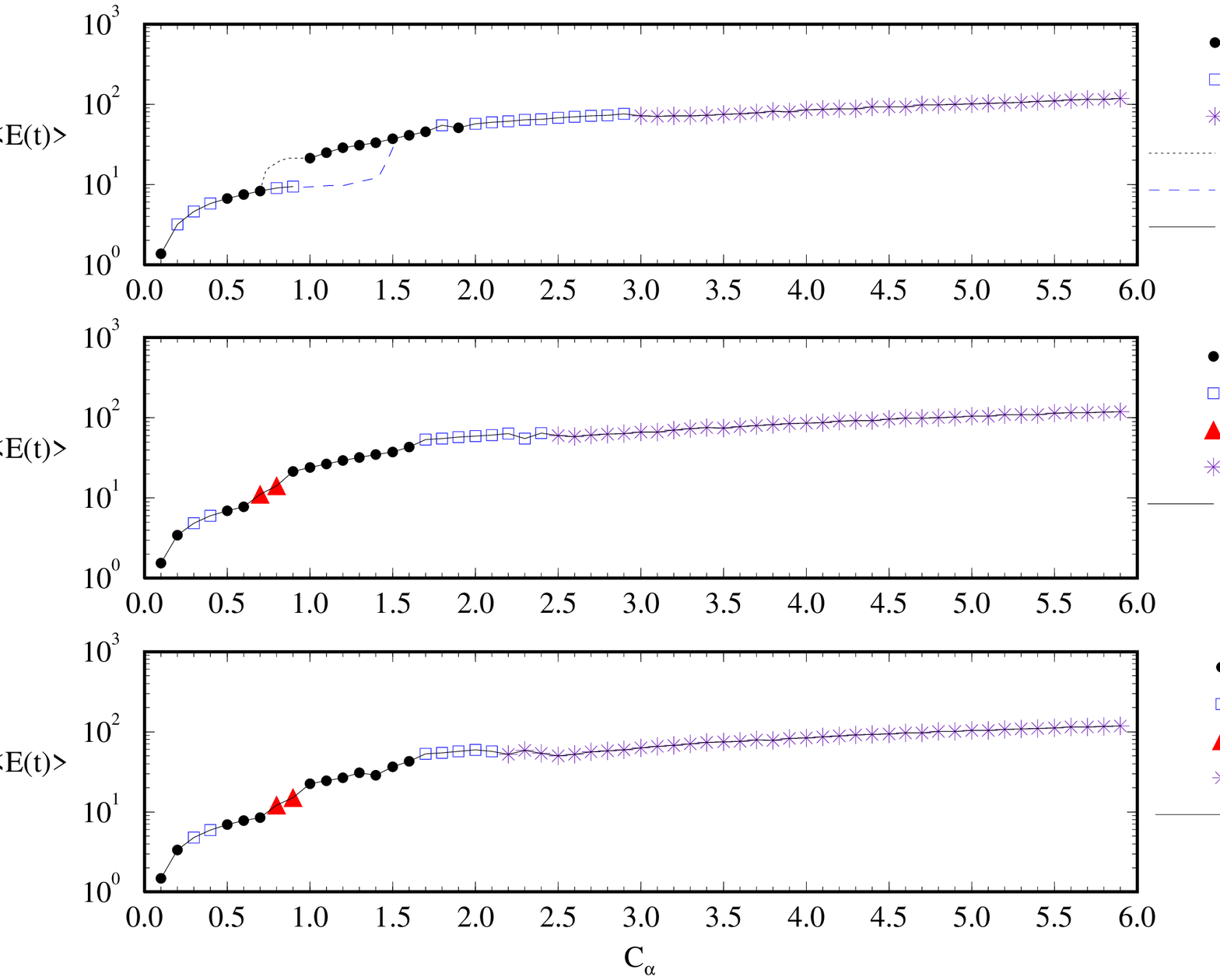}}
\caption[]{\label{mediumstatic}
Energy diagrams for a medium shell with algebraic $\alpha$ dynamo $(r_0=0.5)$
and with $F=0, 0.5$ and $1$. The value of $C_{\omega}$ used was $-10^{5}$. }
\end{figure}

\begin{figure}
\centerline{
\def\epsfsize#1#2{0.5#1}\epsffile{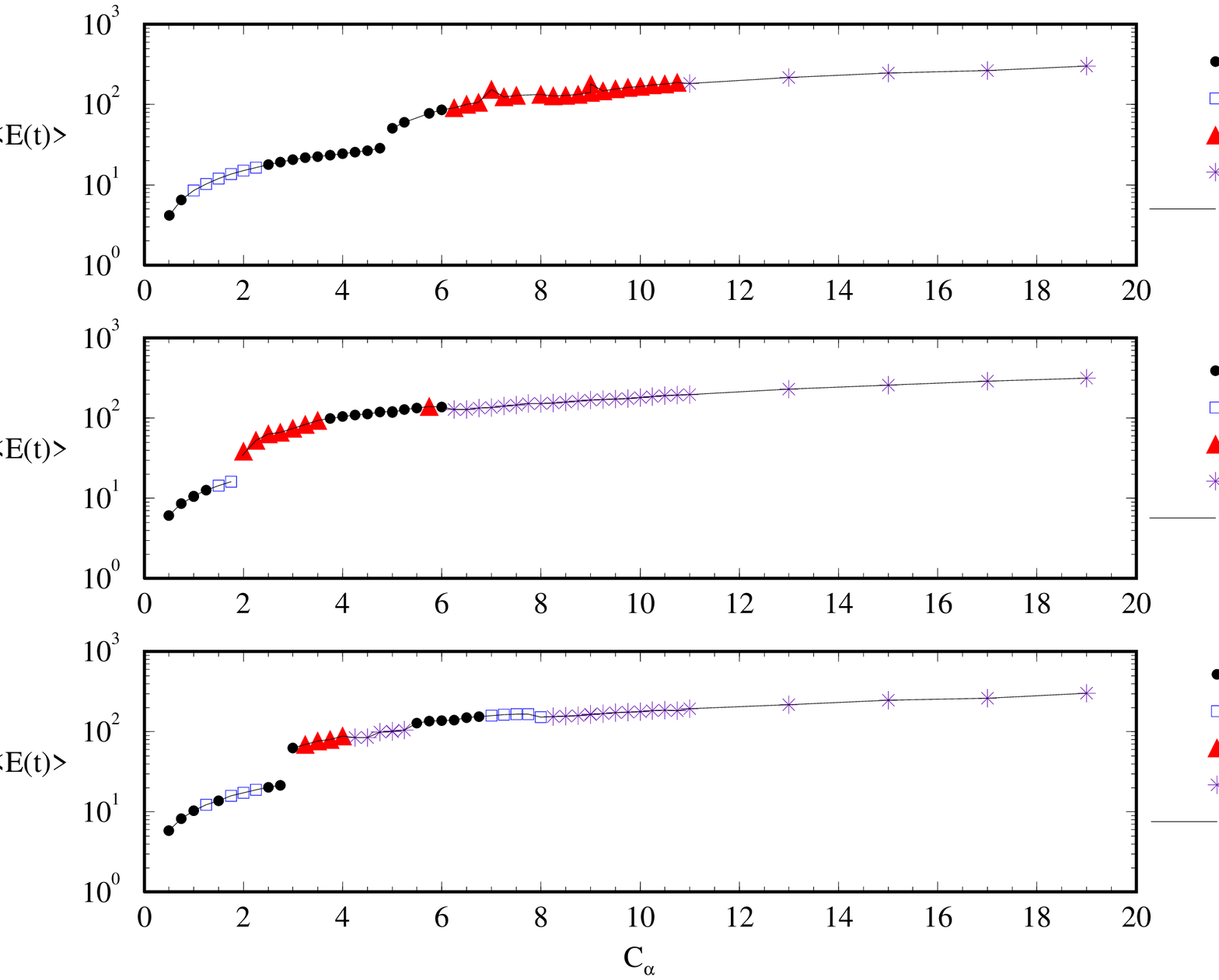}}
\caption[]{\label{thinstatic}
Energy diagrams for a thin shell with algebraic $\alpha$ dynamo
$(r_0=0.7)$ and with $F=0, 0.5$ and $1$. The value of $C_{\omega}$ used was $-10^{5}$. }
\end{figure}

The results of our computations for the corresponding models with
algebraic $\alpha$--quenching, for $C_{\omega} =-10^5$,
are shown in Figs.\ \ref{thickstatic}--\ref{thinstatic} respectively.
%______________________________________________________________________
%______________________________________________________________________
%______________________________________________________________________
%______________________________________________________________________
As can be seen, there are important differences to the models
with dynamic $\alpha$--quenching. The crucial difference
being the very much enhanced likelihood of complicated
(chaotic) modes
of behaviour in
these models, and this
increases with the decreasing thickness
of the shells.
Furthermore, models with $F=0$ are again
somewhat special, because of the occurrence of jumps in energy levels,
which indicates the possibility of hysteresis.
Also, the thick shell case ($r_0=0.2$) is more diverse in its
transition patterns with the
uniformity of the dynamical modes of behaviour growing
as the shell becomes thinner.

To summarise then, comparing the shell models
with the algebraic and dynamic $\alpha$--quenchings,
we observe that there are again important qualitative and quantitative
differences. The introduction of a
dynamic $\alpha$ can have different effects depending
upon the region of the parameter space considered. In particular,
at $C_{\omega}=-10^4$ the dynamic $\alpha$ models
show more variety. However, when comparing models with numerically
allowed upper limits of $C_{\omega}$ the case of algebraic
$\alpha$--quenching can exhibit more complicated modes of behaviour.
%______________________________________________________________________
\section{Multiple attractor regimes}
\begin{figure}
%\picplace{14cm}
\centerline{\def\epsfsize#1#2{0.5#1}\epsffile{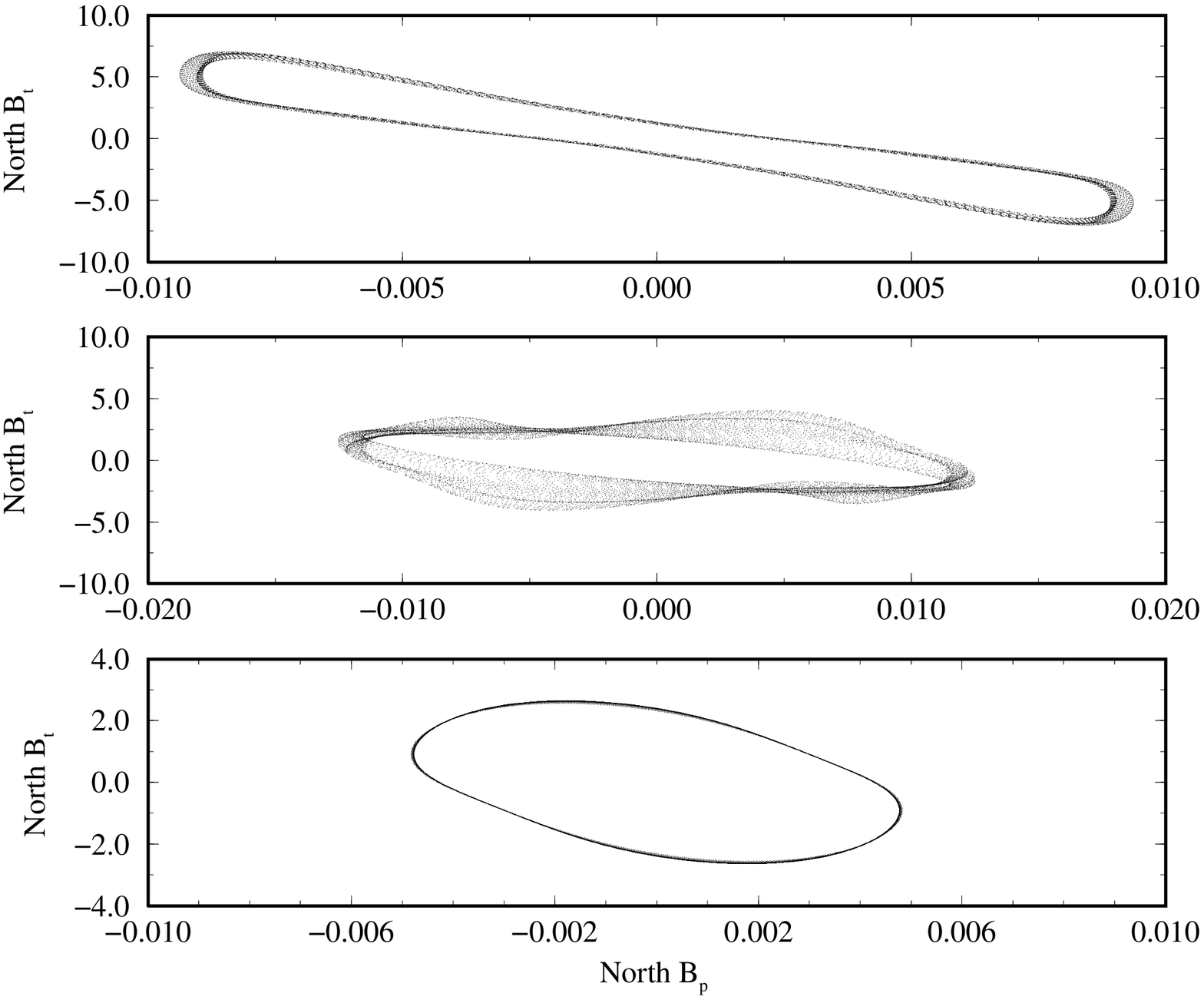}}
\centerline{\def\epsfsize#1#2{0.5#1}\epsffile{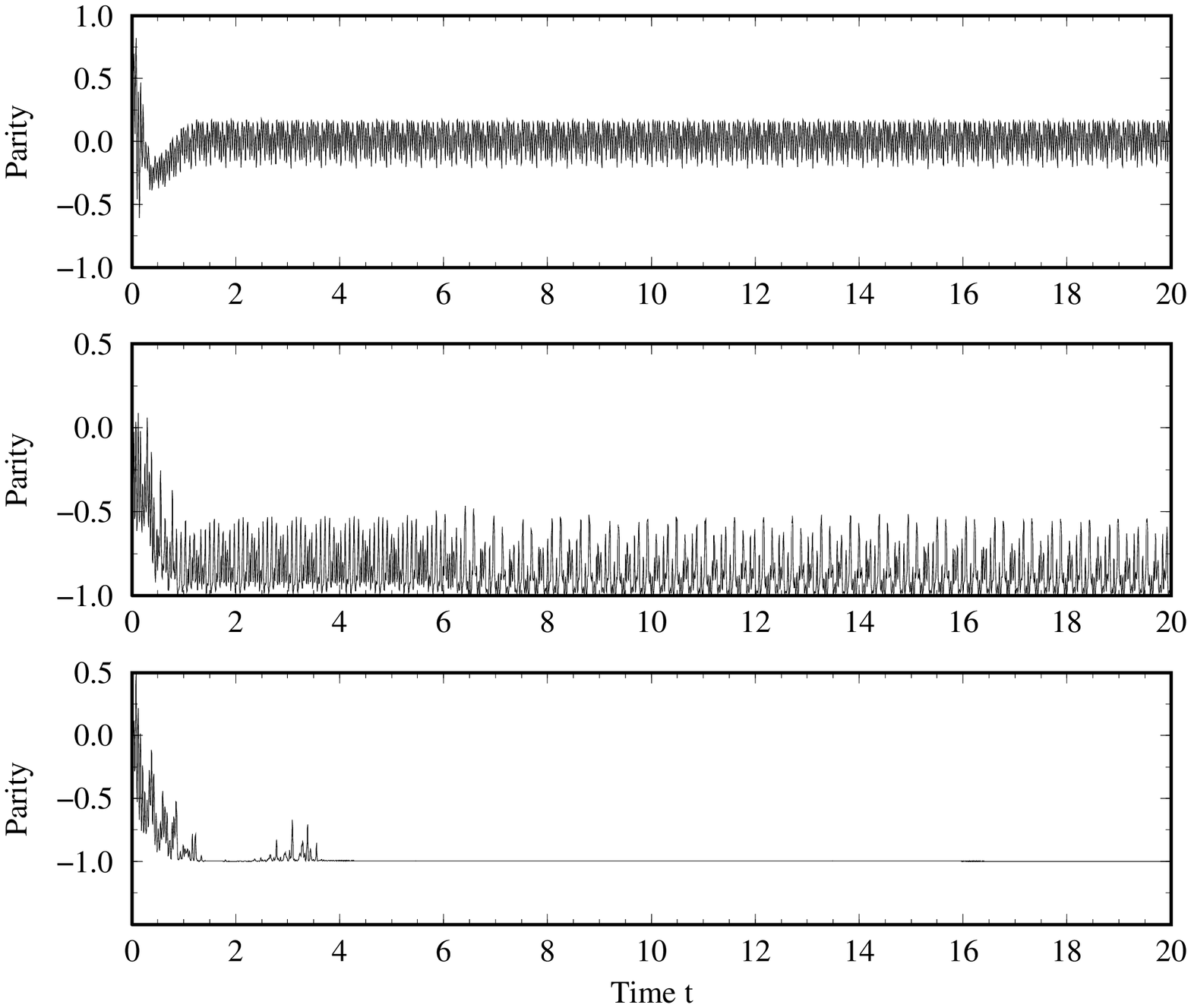}}
\caption[]{\label{timeseries}
Phase space projections and time series of the three (6 due
to the North-South symmetry) types of attractors found
for the dynamic $\alpha$ model with $r_0=0.5$ and $C_{\alpha}=14.6$
and $C_{\omega}=-10^{4}$. For clarity, the projections are drawn
as dots, rather as continuous lines. $B_p$ and $B_t$ are
respectively the poloidal and toroidal parts of the magnetic field
at $r_0=0.6875$ and $\pm 83^\circ$ latitude.}
\end{figure}

%______________________________________________________________________
An interesting feature of the models with
dynamic $\alpha$ is the presence
of parameter intervals over which the system possesses
multiple attractors, ie the occurrence of different dynamical
solutions
possessing different energies at the same value of the control
parameter $C_{\alpha}$, but depending upon the different
initial parities, $P_i$, chosen.
Multiple attractors have previously
been found for axisymmetric two-dimensional dynamo models both with
$\alpha$-quenching (Brandenburg et al.\ 1989b) and with feedback from
the large scale motions (Muhli et al.\ 1995). Tworkowski et al.\ (1997)
have also found evidence for the existence of multiple
attractors, with algebraic $\alpha$--quenching,
but that the types of attractors are different
and the likelihood seems to be less\footnote{We do not distinguish
within chaotic signals those that may be interpreted as
slightly intermittent.}.
Here we
found this mode of behaviour to occur in the intermediate shell models
($r_0=0.5$) in the neighbourhood of $C_{\alpha} \approx 15$ (see
Fig. \ref{medium}). To demonstrate this phenomenon more
clearly, we have plotted in  Fig.\ \ref{timeseries}
two dimensional phase space projections of different solutions,
with the different attractors
being distinguished by their phase space locations.
This distinction is further amplified by comparing the corresponding plots
of the parity, also shown in this figure, which shows three
different types of behaviour. Despite the appearance of behaviours
depicted in this figure, the attractors are nonetheless periodic.

\begin{figure}
%\picplace{7cm}
\centerline{\def\epsfsize#1#2{0.5#1}\epsffile{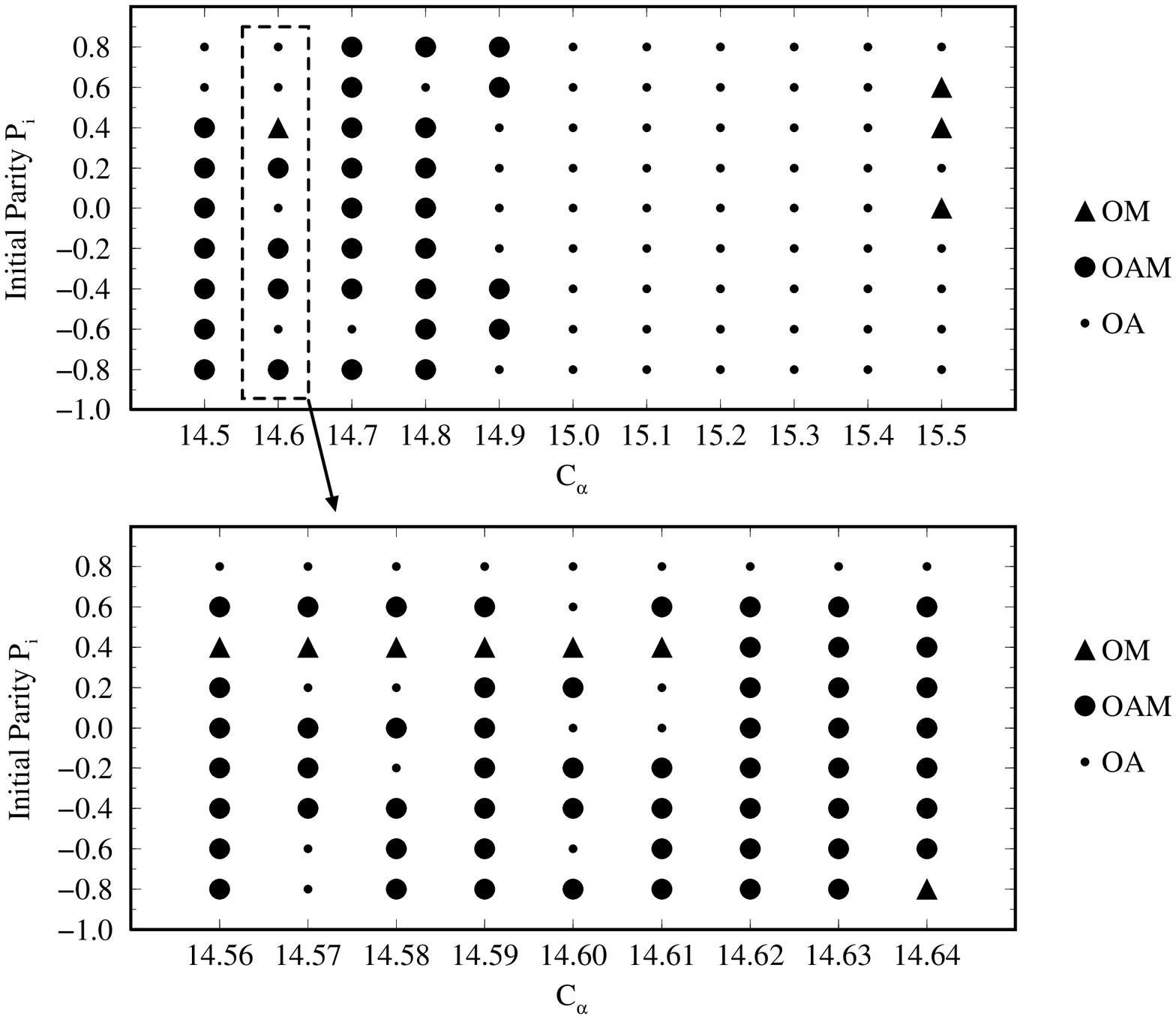}}
\caption[]{\label{fragility}
Fragility with respect to initial parity $P_i$ and initial $C_{\alpha}$.
The symbols A, OAM and OM mean respectively: a periodic (in energy)
pure antisymmetric solution; a periodic (both in energy and parity)
almost pure antisymmetric solution and a periodic (both in energy
and parity) mixed parity solution. }
\end{figure}

Another issue of interest is how sensitive the behaviour of such
models is
with respect to changes in the control parameters
of the system, namely $C_{\alpha}$ and the initial parity $P_i$
(which can be treated as
an initial condition). To study this, we looked at the effects
of changing these parameters
on the
asymptotic properties of the dynamo models.
The results of our calculations are
shown in Fig.\ \ref{fragility}. As can be seen, there is a
region of $C_{\alpha} - P_i$ space in which small changes
can drastically qualitatively change the solution.
Such final state sensitivity (fragility)
was also found in (Tavakol et al.\ 1995) for analogous
models with algebraic $\alpha$--quenching
and also in a more precise
way in finite mode truncated models considered
by Covas and Tavakol (1997), where it was shown
that the basins of attraction
were in fact fractal.
%______________________________________________________________________
%______________________________________________________________________
%______________________________________________________________________
\section{Intermittency}
\begin{figure}
%\picplace{7cm}
\centerline{\def\epsfsize#1#2{0.5#1}\epsffile{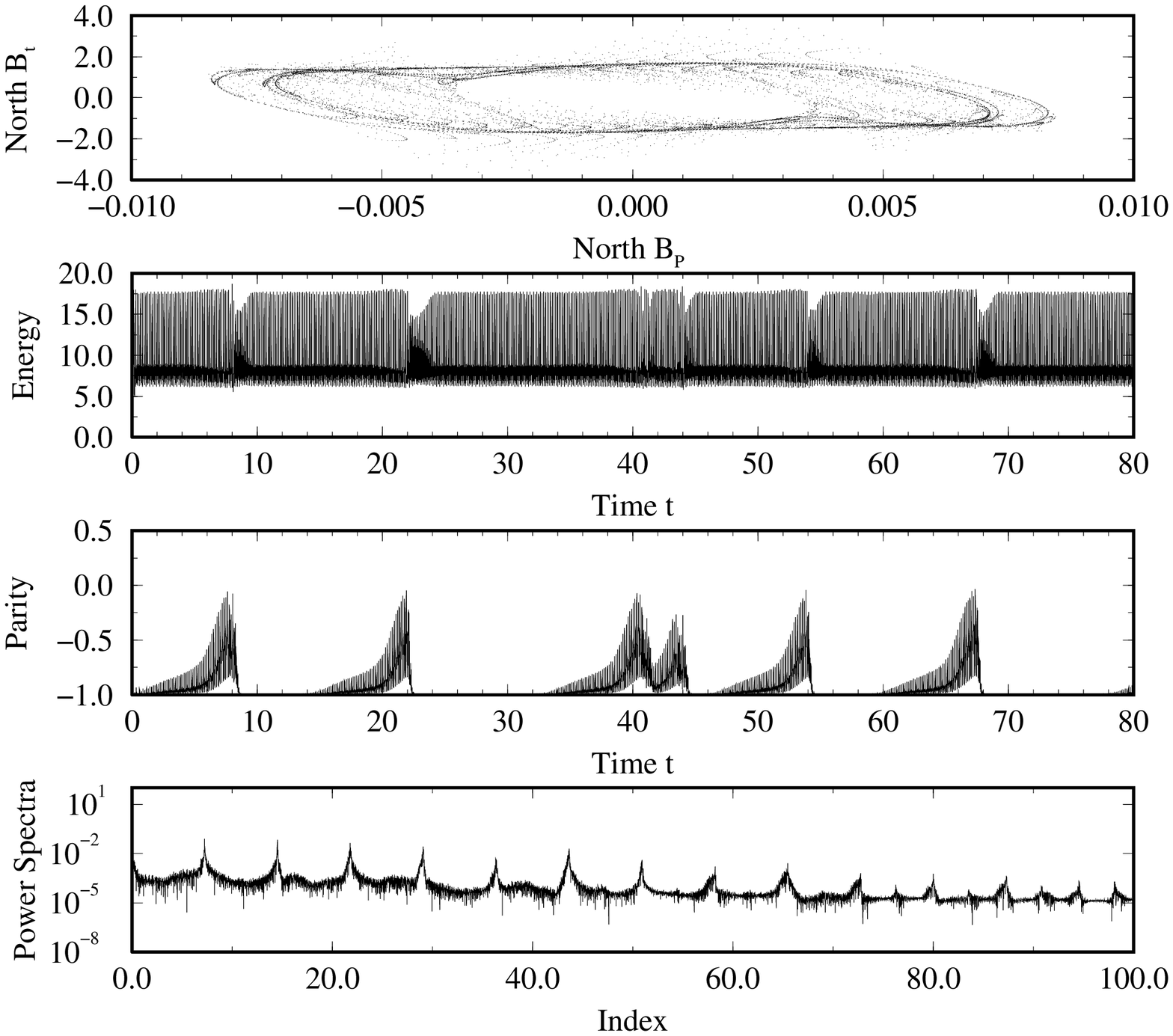}}
\caption[]{\label{kitchatinov}
Result for $r_0=0.5$ with the dynamic $\alpha$ model and a
particular form of $\alpha_h$ (see text).
The parameters are $C_{\alpha}=9.35$, $C_{\omega}=-10^4$ and $F=0$.}
\end{figure}

\begin{figure}
%\picplace{7cm}
\centerline{\def\epsfsize#1#2{0.5#1}\epsffile{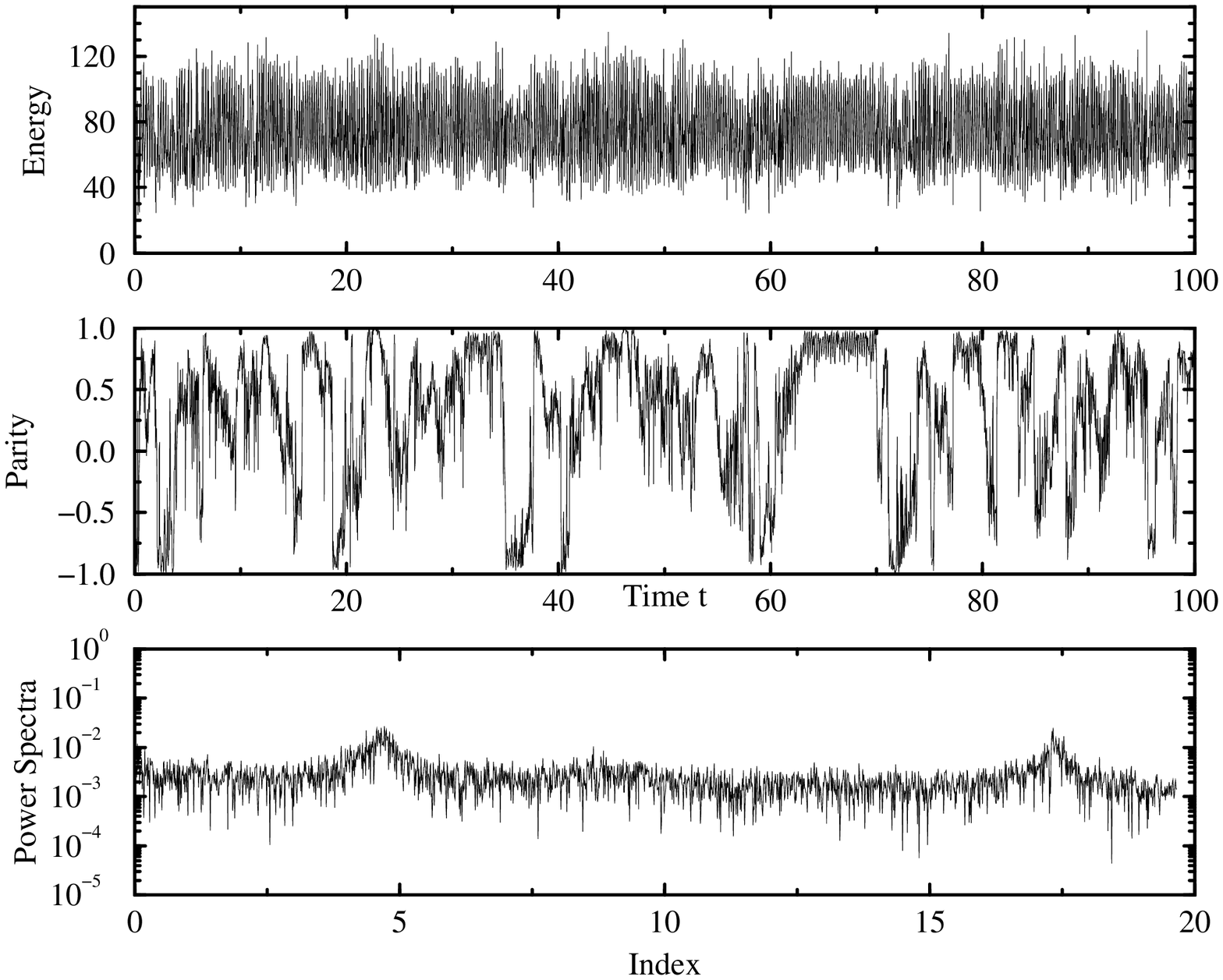}}
\caption[]{\label{chaos}
Result for a full sphere model with an algebraic $\alpha$--quenching.
The parameters are $C_{\alpha}=5.0$, $C_{\omega}=-10^5$ and $F=0$.}
\end{figure}

%______________________________________________________________________
As was mentioned in section \ref{intro}, spherical shell dynamo models
with algebraic $\alpha$--quenching have been shown to be capable of
producing  various  forms of intermittent-type
of behaviour (Tworkowski  et al.\ 1997), ie dynamical modes of
behaviour for which the statistics taken over different time
intervals are different. Now given that the
presence of a dynamic $\alpha$ may drastically reduce the likelihood of
complicated behaviour, the question arises as to whether one
can still have intermittent-type behaviour for models with
dynamic $\alpha$. We found that the presence of dynamic $\alpha$--quenching
greatly reduced this possibility and
despite the relatively extensive numerical
studies of these models, reported above,
the only examples of this type of behaviour that we were able
to find were in a shell model with $r_0=0.5$
in which the
algebraic part of the $\alpha$--quenching
was taken to be the form given in Kitchatinov (1987).
Fig.\ \ref{kitchatinov} shows an example
of such a behaviour, whereby intervals of parity being
very nearly antisymmetric are punctuated by migrations
towards zero parity and then a sudden drop. This mode of
behaviour is an example of the type of intermittency referred
to as {\it icicle} intermittency (Brooke and Moss 1995, Brooke 1997,
Brooke et al.\ 1997). To distinguish between this intermittent
mode of behaviour with
chaotic behaviour, we show in Fig.\ \ref{chaos} an example
of the latter kind.

The point here is that overall, the employment of a dynamic $\alpha$
seems to decrease the likelihood of intermittent-type behaviour
relative to the cases where algebraic $\alpha$--quenching is used.
%______________________________________________________________________
%______________________________________________________________________

%______________________________________________________________________
%______________________________________________________________________
%______________________________________________________________________
\section{Conclusion}
%______________________________________________________________________
We have made a study of axisymmetric
mean field spherical and spherical shell dynamo models,
with both dynamic and algebraic $\alpha$--quenchings.

In the full sphere models, the main similarities are
the occurrence of symmetric, antisymmetric,
mixed modes and absence of intermittent modes of behaviour.
The differences are in the details of the transitions
between the different modes of behaviour and, more significantly,
the presence of chaotic behaviour in the algebraic
$\alpha$--quenching case. As far as we are aware,
this is the first time chaotic behaviour
has been observed in full sphere models of this type.

For the spherical shell models, we again observe important qualitative
and quantitative differences. In particular, the effect of introducing
dynamic $\alpha$--quenching depends crucially on the region of the
parameter space considered. For $C_{\omega}=-10^4$, the dynamic $\alpha$
models seem to produce more varied modes of behaviour. On the other
hand, considering the numerical upper bound of $C_{\omega}$ in each case
shows that the introduction of dynamic $\alpha$--quenching drastically
reduces the likelihood of the occurrence of chaotic behaviour, which was
observed in models with algebraic $\alpha$--quenching.  These models
also show multi-attractor regimes with the possibility of the final
state sensitivity (fragility) with respect to small changes in
$C_{\alpha}$ and the initial parity, as well as intermittent modes of
behaviour.  We also observe that, in the highly nonlinear regimes (with
the extreme $C_{\omega}$ values), the symmetric modes are preferred in
the full sphere models and thick shells while the reverse appears to be
the case for intermediate and thinner shells.  This is in contrast to
kinematic theory, where the most preferred mode is antisymmetric for the
full sphere and thick shell cases, and symmetric for thinner shells
(Roberts 1972).

It might have been expected that since making the $\alpha$ effect
dynamical adds more complexity to the system, that it should have
uniformly enhanced the complexity of its possible modes of behaviour.
Indeed previous experiments with a different form of time dependent
$\alpha$ suggest that the system could then exhibit more complicated
temporal behaviour.  For example, Yoshimura (1978) investigated systems
with a time-delay built into the $\alpha$ effect.  Our present results
show that the picture is likely to be complicated with the outcome
depending on the region of parameter space considered, rather than a
simplistic decrease or increase in complexity.  To establish the true
measure of this change requires an extensive study of the behaviour of
these models in the $C_{\alpha}$--$C_{\omega}$ parameter space, to which
we shall return in a future publication.

Finally, in view of the fact that the derivation of an $\alpha$ with
dynamic quenching is more general (because it allows for explicit
time-dependence), the results thus obtained are more realistic than
those obtained using the algebraic form.

%______________________________________________________________________
\begin{acknowledgements} EC is supported by grant  BD / 5708 / 95 --
Program PRAXIS XXI, from JNICT -- Portugal.
RT benefited from SERC UK Grant No. H09454. This research also benefited
from the EC Human Capital and Mobility (Networks) grant ``Late type stars:
activity, magnetism, turbulence'' No. ERBCHRXCT940483.
\end{acknowledgements}

%-----------------------------------------------------------------

%______________________________________________________________________

%______________________________________________________________________
\end{document}